\documentclass[a4paper,fleqn,usenatbib]{mnras}

\usepackage[T1]{fontenc}
\usepackage{ae,aecompl}

\usepackage{longtable}
\usepackage{graphicx}	
\usepackage{amsmath}	
\usepackage{amssymb}	

\usepackage{pdflscape}
\usepackage{multicol}
\usepackage{multirow}

\title[The symbiotic nature of SMP LMC 88]{A D'-type symbiotic binary in the planetary nebula SMP\,LMC\,88\thanks{Based on observations made with the Southern African Large Telescope (SALT) under  programmes 2015-2-SCI-036 and 2017-1-SCI-049}}
\author[I{\l}kiewicz et al.]{Krystian I{\l}kiewicz,$^{1}$\thanks{E-mail: ilkiewicz@camk.edu.pl} Joanna Miko{\l}ajewska,$^{1}$  Brent Miszalski,$^{2,3}$ \newauthor Szymon Koz{\l}owski$^{4}$ and Andrzej Udalski$^{4}$\\
$^{1}$Nicolaus Copernicus Astronomical Center, Polish Academy of Sciences, ul. Bartycka 18, 00-716 Warsaw, Poland \\
$^{2}$South African Astronomical Observatory, PO Box 9, Observatory, 7935, South Africa \\
$^{3}$Southern African Large Telescope Foundation, PO Box 9, Observatory, 7935, South Africa \\
$^{4}$Warsaw University Observatory, Al. Ujazdowskie 4, 00-478 Warszawa, Poland}

\date{Accepted XXX. Received YYY; in original form ZZZ}

\pubyear{2017}

\begin{document}
\label{firstpage}
\pagerange{\pageref{firstpage}--\pageref{lastpage}}
\maketitle

\begin{abstract}
SMP LMC 88 is one of the planetary nebulae (PN) in the Large Magellanic Cloud. We identify in its spectrum Raman scattered \mbox{O\,{\sc vi}} lines at 6825 and 7083{\AA}. This unambiguously classifies the central object of the nebula as a symbiotic star (SySt). We identified the cold component to be a K-type giant, making this the first D'-type (yellow) SySt discovered outside the Galaxy. The photometric variability in SMP LMC 88 resembles the the orbital variability of Galactic D'-type SySt with its low amplitude and sinusoidal lightcurve shape. The SySt classification is also supported by the He I diagnostic diagram.
 
\end{abstract}





\section{Introduction}

Symbiotic stars (SySt) are binary systems consisting of a red giant (RG) transferring matter to a much smaller, hot companion. Most of the SySt host a white dwarf (WD) as a compact companion, but sometimes a neutron star is observed. S-type (stellar) SySt host a normal red giant, while in D-type SySt a Mira surrounded by a dense nebula is present. In the case of a donor with an earlier (F--K) spectral type the system is classified as yellow SySt. D'-type (dusty) SySt are yellow SySt which additionally shows emission from dust in infrared (IR). SySt can exhibit a variety of phenomena, such as active phases, nova outbursts or jets. Given their complex interactions they are good laboratories to study evolution and variability of stars in binary systems. A recent review of SySt is presented in \citet{2012BaltA..21....5M}.

In over a dozen SySt optical nebulae have been resolved. Among the best studied symbiotic nebulae are the ones in R~Aqr \citep{1985A&A...148..274S}, HM~Sge \citep{1984A&A...139..296S}, Hen~2-104 \citep{2008A&A...485..117S},  Hen~2-147 \citep{2007A&A...465..481S}, HBV~475 \citep{1997A&A...324..606S} and V1016 Cyg \citep{1999A&A...348..978C}. The sizes of symbiotic nebulae range from $\sim500$~AU \citep[HBV~475][]{1997A&A...324..606S} up to $\sim$2~pc \citep[V417 Cen][]{1994A&A...285..241V} in the optical, but they can be several times larger in the IR \citep{2008ApJ...682.1087M}. The nebulae are typically very young, with kinematic ages ranging from $\sim200$~yrs \citep[Hen 2-147;][]{2007A&A...465..481S} up to $\sim3000$~yrs \citep[BI Cru;][]{1992A&A...265L..37S}. Moreover, they can show a variety of shapes, including multiple shells  \citep[e.g. Hen 2-104][]{2008A&A...485..117S}. The review of their properties is presented in \citep{1997ppsb.conf..147C}.

From a sample of 71 SySt \citet{1999A&A...343..841C} concluded that resolved optical nebulae are present in $\sim40\%$ of D-type SySt, and are very rare around S-type SySt. Almost all nebulae around SySt are not genuine PNe, i.e. asymptotic giant branch (AGB) remnants \citep{1999A&A...343..841C,2003ASPC..303..428K}, but just nova shells or wind lost by the mass donor which was ionized by a WD. This results from the fact that in D-type SySt the mass donor is a thermally pulsating AGB star (Mira), which is a relatively short phase of stellar evolution just before turning into AGB remnant. If the nebula would come from the current WD, this would imply a small difference between initial masses of the system components, which would be extremely unlikely for such a big number of D-type SySt. The nebulae around D-type systems most probably consists of a material lost by the present Mira star which may be overtaken and accelerated by nova explosion of the WD \citep{1997ppsb.conf..147C}.

A separate class of objects are nebulae around yellow D'-type SySt. Yellow systems are very rare in general \citep{1991A&A...243..469S}, and only in two cases optical nebula have been resolved around bona fide D'-type SySt, AS~201 \citep{1991A&A...243..469S} and  V417~Cen \citep{1994A&A...285..241V}. Moreover, connections with SySt have been suggested for other objects with resolved nebulae \citep[e.g.][]{2012MNRAS.419...39M}. Nebulae around bona fide D'-type SySt seem older than in the case of D-type systems, namely 4000~yrs for AS~201 \citep{1991A&A...243..469S} and 40000~yrs for V417~Cen \citep{1994A&A...285..241V}. Their expansion velocities are of order of ten km/s, similar to terminal velocities of Mira winds and an order of magnitude lower than in the case of nebulae around D-type SySt \citep{1997ppsb.conf..147C}. The same statistical argument about origin of the nebula as in the case of nebulae around D-type SySt can not be used in the case of D'-type systems, since they do not host a Mira, but an earlier type giant. Moreover, the early type giant has a significantly weaker wind with a low mass loss rate. This wind cannot provide enough material for a shell, as opposed to the D-type SySt with a Mira mass donor. Overall, it seems that nebulae around D'-type SySt are envelopes ejected by a former AGB star, which is currently a WD and a mass acceptor in a SySt \citep{1991A&A...243..469S,1993A&A...277..195M}. This means the these systems are simultaneously SySt as well as genuine PNe.

In this work we present a study of SMP~LMC~88, which have been classified as a PN in the Large Magellanic Cloud (LMC) \citep{1978PASP...90..621S,1984MNRAS.208..633M}. Based on a discrepancy between derived properties of the nebula from different models, \citet{1991ApJ...367..115D} suggested that the central star of SMP~LMC~88 is a binary based on the big difference between derived Zanstra-temperature and the excitation balance temperature. \citet{1996A&A...307..359K} suggested that this object could be associated with a globular cluster 1280~HS398. The nebula around the object is elliptical with faint emission around a bright core \citep{2006ApJS..167..201S}. The chemistry of the nebula is somewhat unusual for a PN, including low oxygen content and lack of detection of carbon, which suggest that the PN originates from a $\sim$3~M$_\odot$ star \citep{2015MNRAS.452.3679V}. 

We present our observations of SMP~LMC~88 in Section~\ref{obs_sec}. The classification of the system is presented in Section~\ref{class_sec}. The properties of the nebular around SMP~LMC~88 are calculated and discussed in Section~\ref{neb_sec}. Search for variability of the object is presented in Section~\ref{var_sec}.

\section{Observations}\label{obs_sec}

In our study we used a spectrum taken with the AAOmega spectrograph on the Anglo-Australian Telescope (AAT) with exposure time of 1.5hr from \citet{2013ApJ...775...92K}, where the the details of the instrument setup and reduction process are described. Additionally, we collected emission line fluxes from the literature. The emission line fluxes are presented in Table~\ref{table:lineratios}.  

\begin{figure*}
  \resizebox{\hsize}{!}{\includegraphics{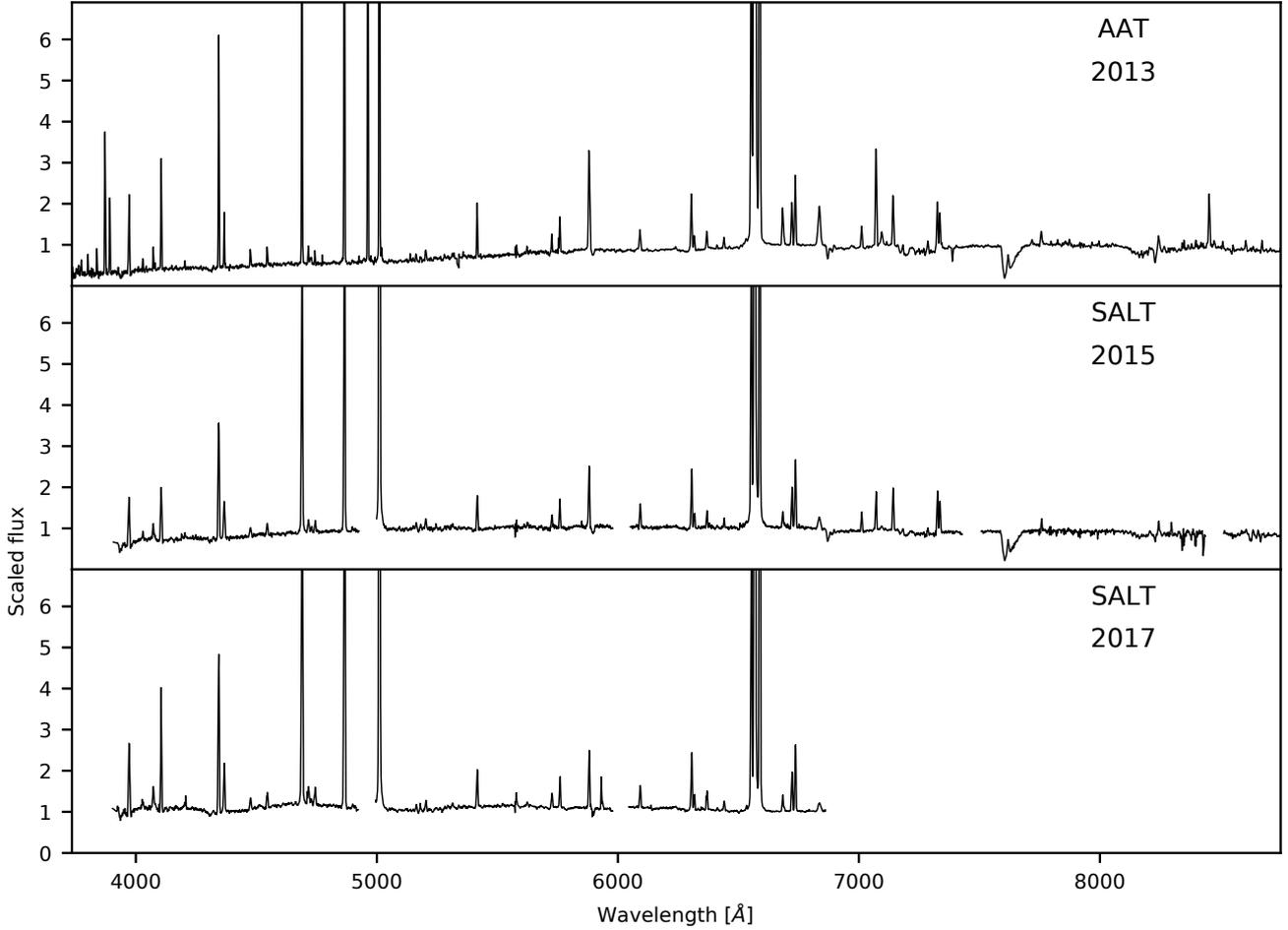}}
  \caption{The AAT and SALT spectra of SMP~LMC~88. }
  \label{SMP88_aaomega}
\end{figure*}

We carried out two observations of SMP~LMC~88 (RA=05:42:33.193 DEC=-70:29:24.08) with the Robert Stobie Spectrograph \citep{2003SPIE.4841.1463B,2003SPIE.4841.1634K} mounted on the Southern African Large Telescope \citep[SALT;][]{2006MNRAS.372..151O}. The spectra were carried out with PG0900 grating and a slit width of 1.5~arcsec, resulting in a resolution of R$\simeq$1000. In 2015 two 2000s exposures were taken under programme 2015-2-SCI-036 (PI: Miko{\l}ajewska) using two grating angles in order to cover a combined spectral range of 3930--9300\AA . In 2017 one spectrum was taken with a 2700s exposure time under programme 2017-1-SCI-049 (PI: I{\l}kiewicz), covering only the blue part of spectrum. The reduction was performed with the \textsc{pysalt} package  \citep{2010SPIE.7737E..25C} and standard \textsc{IRAF} procedures. The measured errors in the line fluxes are mostly related to uncertainty in choosing the continuum level. The errors are 15\% for fluxes of the strongest lines and 30\% for the weakest lines. The AAT and SALT spectra are presented in Fig.~\ref{SMP88_aaomega}.

\begin{figure}
  \resizebox{\hsize}{!}{\includegraphics{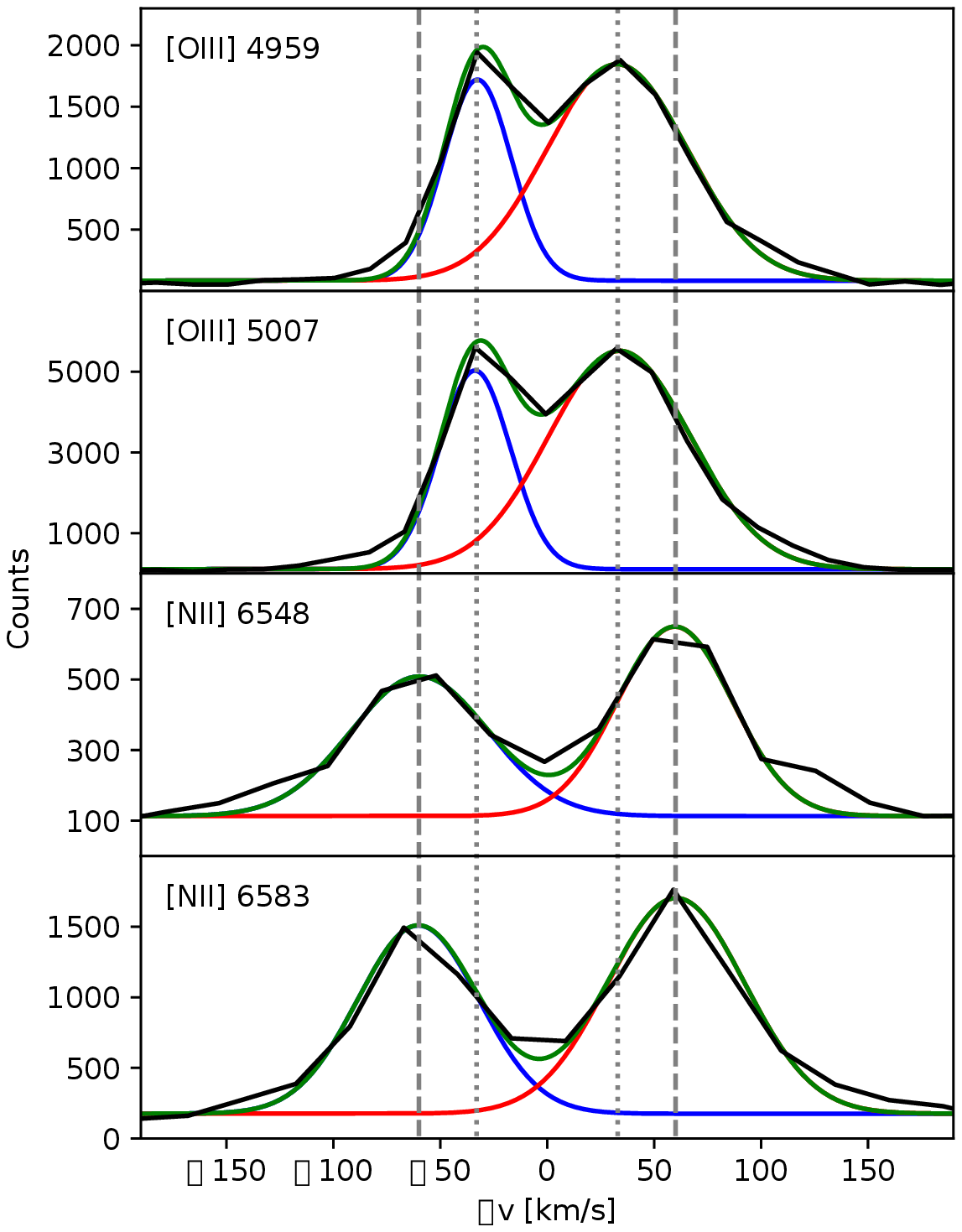}}
  \caption{The radial velocities of emission lines in SMP~LMC~88 on the HST/STIS spectrum. The black line is the observed flux, the blue and red lines are fitted Gaussian components and the green line is the sum of two Gaussian functions. The dotted gray line is the measured radial velocity of two [\mbox{O\,{\sc iii}}] lines, while the dashed gray line is the measured radial velocity of two [\mbox{N\,{\sc ii}}] lines.}
  \label{HST_lines}
\end{figure}

The radial velocities of lines were analyzed with the Hubble Space Telescope (HST) spectra from the Space Telescope Imaging Spectrograph \citep[STIS;][]{1998ApJ...492L..83K,1998PASP..110.1183W}. The spectra were already partially analyzed by \citet{2006ApJS..167..201S}. The emission line profiles are presented in Fig.~\ref{HST_lines}.

\begin{table*}
\caption{Observed emission lines flux ratios of SMP~LMC~88. The fluxes taken from \citet{1991ApJS...75..407M} were originally reddening corrected. We transformed the fluxes back to the observed values using the reddening value adopted by \citet{1991ApJS...75..407M}.}         
\label{table:lineratios}     
\centering                       
\begin{tabular}{ccccccc}       
\hline\hline   
		  \multirow{ 2}{*}{Source}  & \citeauthor{2006AA...456..451L} & \citeauthor{1991ApJS...75..407M}  & \citeauthor{2006ApJS..167..201S} & \multirow{ 2}{*}{AAT} &    \multirow{ 2}{*}{SALT} &    \multirow{ 2}{*}{SALT}\\
		    & \citeyearpar{2006AA...456..451L} & \citeyearpar{1991ApJS...75..407M}  & \citeyearpar{2006ApJS..167..201S} &  &   & \\
		 Date   & 1982--1991 & 1983--1986  & 30.08.2001 & 07.01.2013 &   18.11.2015 & 23.10.2017 \\
		 MJD   & & & &  56299.6 &  57345.0 & 58049.0  \\
\hline

ID	&	  \multicolumn{5}{c}{100$\times$Flux/Flux(H$\beta$)}\\
\hline
H10+\mbox{He\,{\sc ii}}~3798 			&	 	&	3	&	 	&		3		&				&		\\
H9+\mbox{He\,{\sc ii}}~3835		&	 	&	4	&	   	&		4		&				&		\\
$[$\mbox{Ne\,{\sc iii}}]~3869		&	 39.4 	&	27	&	  	&		21		&				&		\\
H8+\mbox{He\,{\sc i}}+\mbox{He\,{\sc ii}}~3889		&	 16.7 	&	14	&	 	&		12		&				&		\\
$[$\mbox{Ne\,{\sc iii}}]~3969+H7		&	 19.2 	&	15	&	  	&		16		&		14		&	18	\\
\mbox{He\,{\sc i}}+\mbox{He\,{\sc ii}}~4026		&	 	&	3	&	  	&		2		&		2		&	2	\\
$[$\mbox{S\,{\sc ii}}]~4070		&	 	&	10	&	  	&		3		&		4		&	5	\\
H$\delta$		&	 17.5 	&	18	&	 	&		17		&		15		&	23	\\
H$\gamma$		&	 27.2 	&	38	&	  	&	 35		&		34		&	39	\\
$[$\mbox{O\,{\sc iii}}]~4363		&	 7.6 	&	12	&	  	&		8		&		11		&	11	\\
\mbox{He\,{\sc i}}~4471		&	 	&	5	&	 	&	 	3		&		3		&	3	\\
\mbox{He\,{\sc ii}}~4541		&	 1.4 	&	3	&	 	&	 	3		&		3		&	4	\\
\mbox{He\,{\sc ii}}~4686		&	 73.0 	&	84	&	  	&		71		&		84		&	85	\\
$[$\mbox{Ar\,{\sc iv}}]~4712		&	 2.6 	&	5	&	  	&		3		&		4		&	4	\\
$[$\mbox{Ar\,{\sc iv}}]~4741		&	 3.2 	&	7	&	  	&				&		3		&	4	\\
H$\beta$		&	100	&	100	&	100	&	 100 		&		100		&	100	\\
$[$\mbox{O\,{\sc iii}}]~4959		&	131	&	94	&	  113.5 	&		88		&				&		\\
$[$\mbox{O\,{\sc iii}}]~5006		&	402	&	272	&	 333.0 	&		268		&		425		&	364	\\
\mbox{He\,{\sc ii}}~5412		&	 8.6 	&	6	&	 	&	 	9		&		8		&	8	\\
$[$\mbox{N\,{\sc ii}}]~5755		&	 10.4 	&	2	&	 	&		7		&		6		&	6	\\
\mbox{He\,{\sc i}}~5876		&	 17.4 	&	12	&	 	&	 	24		&		15		&	12	\\
$[$\mbox{Ca\,{\sc v}}]+$[$\mbox{Fe\,{\sc vii}}]~6088		&	 	&	  	&	 	&		5		&		6		&	5	\\
$[$\mbox{O\,{\sc i}}]~6302		&	 22.4 	&	10	&	  9.8 	&		14		&		14		&	13	\\
$[$\mbox{S\,{\sc iii}}]+\mbox{He\,{\sc ii}}~6312		&	 4.0 	&	 	&	  3.5 	&		4		&		5		&	3	\\
$[$\mbox{O\,{\sc i}}]~6365		&	 7.0 	&	5	&	  3.7 	&		4		&		4		&	4	\\
$[$\mbox{Ar\,{\sc v}}]~6436		&	 3.4 	&	 	&	  	&		3		&		2		&	2	\\
$[$\mbox{N\,{\sc ii}}]~6548		&	110	&	46	&	  50.0 	&		88		&		85		&	69	\\
H$\alpha$		&	685	&	455	&	  453.8 	&		895		&		586		&	452	\\
$[$\mbox{N\,{\sc ii}}]~6584		&	314	&	139	&	  155.2 	&		219		&		234		&	194	\\
\mbox{He\,{\sc i}}+\mbox{He\,{\sc ii}}~6678		&	 6.6 	&	8	&	  5.8 	&		10		&		5		&	4	\\
$[$\mbox{S\,{\sc ii}}]~6716		&	 15.0 	&	5	&	  7.9 	&		10		&		10		&	8	\\
$[$\mbox{S\,{\sc ii}}]~6730		&	 23.8 	&	8	&	  13.1 	&		17		&		17		&	14	\\
\mbox{O\,{\sc vi}}~6825		&	 	&	 	&	 	&	 	18		&		6		&	4	\\
$[$\mbox{Ar\,{\sc v}}]~7007		&	 7.7 	&	 	&	  	&		6		&		5		&		\\
\mbox{He\,{\sc i}}~7065		&	 14.3 	&	9	&	 	&	 	24		&		10		&		\\
\mbox{O\,{\sc vi}}~7083		&	 	&	 	&	 	&	 	6		&		1		&		\\
$[$\mbox{Ar\,{\sc iii}}]~7136		&	 17.2 	&	  	&	 	&		12		&		11		&		\\
$[$\mbox{O\,{\sc ii}}]~7320		&	 15.8 	&	  	&	 	&		11		&		11		&		\\
$[$\mbox{O\,{\sc ii}}]~7330		&	 	&	 	&	  	&		9		&		8		&		\\
$[$\mbox{Ar\,{\sc iii}}]~7751		&	 	&	 	&	  	&		4		&		3		&		\\
		&	 	&	 	&	 	&		11		&				&		\\
$[$\mbox{S\,{\sc iii}}]~9069		&	 	&	 	&	  	&				&		21	   	&		\\

\hline                   
\end{tabular}
\end{table*}

\begin{figure}
  \resizebox{\hsize}{!}{\includegraphics{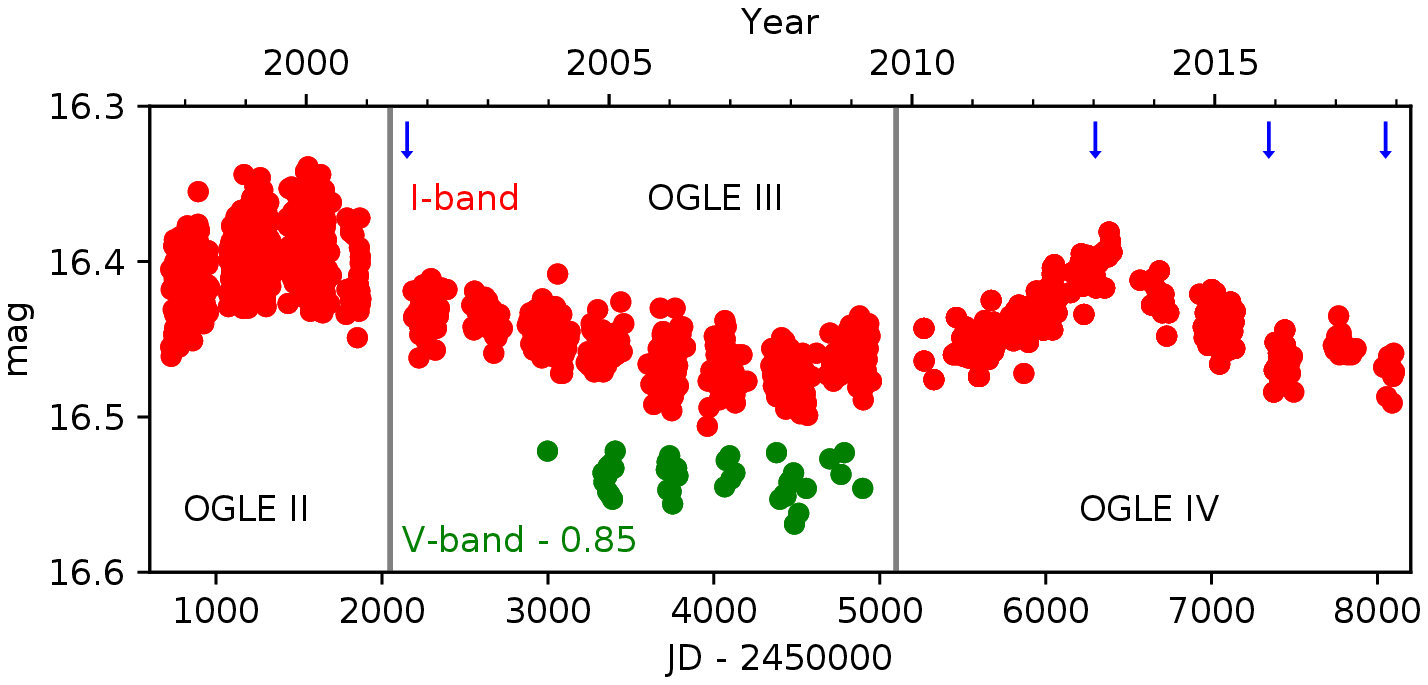}}
  \caption{ $V$- (green) and $I$-band (red) OGLE light curves for SMP LMC 88. The two maxima at approximately JD'=1400 and 6300 are clearly visible, pointing to a plausible period of $\sim$4900 days. The blue arrows indicate dates of spectra from Table~\ref{table:lineratios}. }
  \label{SMP88_ogle}
\end{figure}

The photometric observations presented in this work come from the Optical Gravitational Lensing Experiment \citep[OGLE;][]{2015AcA....65....1U}. The OGLE identification numbers of SMP LMC 88 are: OGLE-II LMC SC18 180289, OGLE-II LMC SC19 28850, OGLE-III LMC177.2 82390, and OGLE-IV LMC552.23 592D. We combined the OGLE-II, OGLE-III, and OGLE-IV light curves for this source, now spanning 18.4 years, and present them in Fig.~\ref{SMP88_ogle}.

\section{Results}

\subsection{Symbiotic classification}\label{class_sec}

\begin{figure}
  \resizebox{\hsize}{!}{\includegraphics{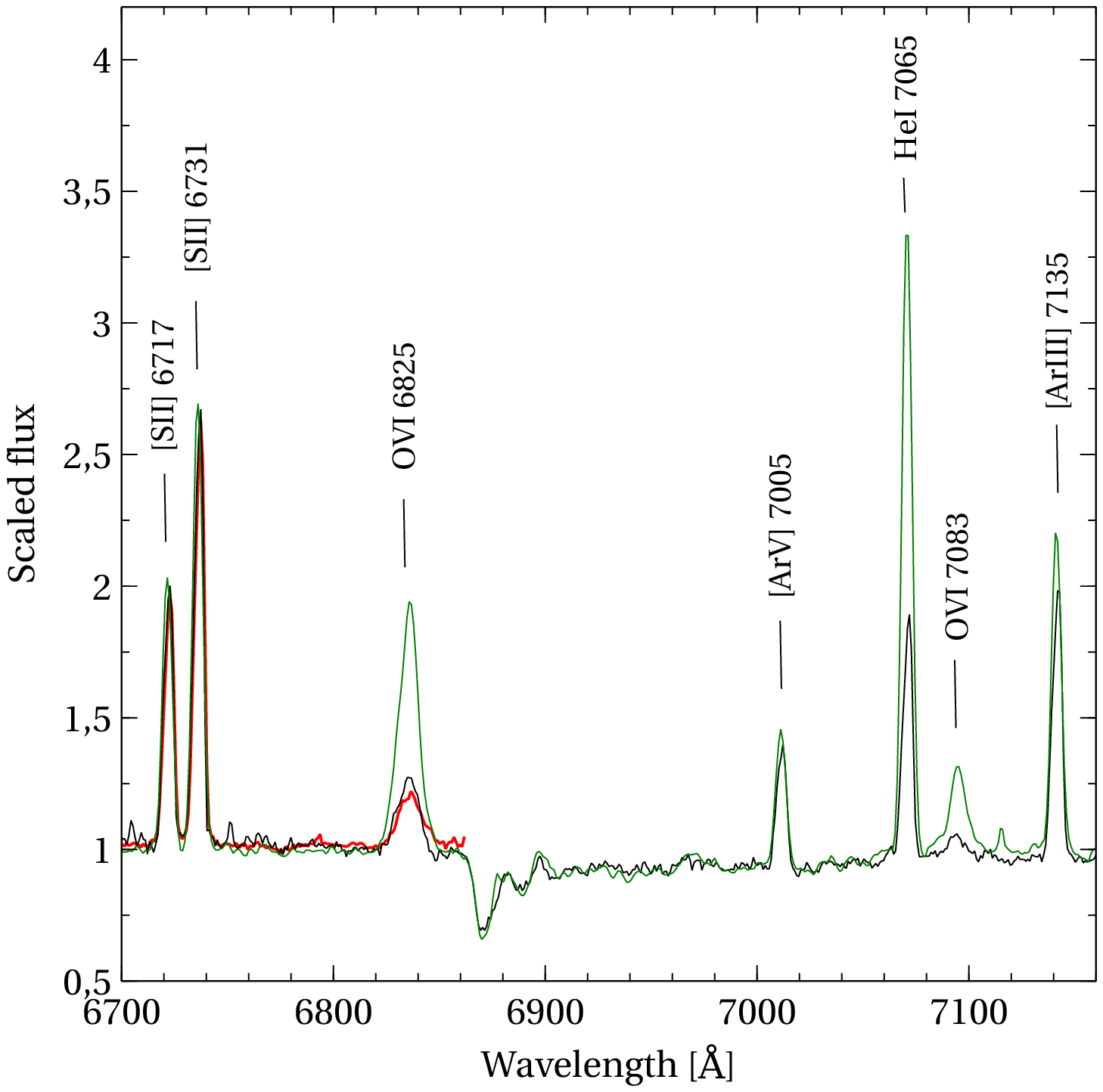}}
  \caption{AAT (green line), 2015 SALT (black line) and 2017 SALT (red line) spectra of SMP~LMC~88 scaled so that the local continuum around 6825 and 7083{\AA} Raman scattered  \mbox{O\,{\sc vi}} lines would be at the same level.}
  \label{SMP88_raman}
\end{figure}

The  SMP~LMC~88 spectra reported by \citet{1996A&AS..116...95L} and \citet{2006AA...456..451L} show an emission line at 6825{\AA} but  the authors were not able to identify it.
 This line is also present in our spectra (Fig.~\ref{SMP88_aaomega}).
We identify this feature as the Raman scattered \mbox{O\,{\sc vi}} line \citep{1989A&A...211L..27N}. The identification is supported by the presence of the accompanying Raman scattered \mbox{O\,{\sc vi}} line at 7083{\AA} (Fig.~\ref{SMP88_raman}). The presence of the 6825 and 7083{\AA} lines unambiguously reclassifies this object as a SySt \mbox{\citep[see e.g.][and references therein]{2014A&A...570L...4S}}. The SySt classification is additionally supported by the fact that thus far only  Raman scattered \mbox{He\,{\sc ii}} lines have been observed in young PNe \citep{1997A&A...323..217P,2001ApJ...551L.121L,2006ApJ...636.1045L,2009ApJ...695..542K}, which we do not detect in our spectra, while the  6825{\AA} line is one of the strongest emission lines in the optical range. Since SMP~LMC~88 shows a resolved nebula \citep[e.g.][]{2006ApJS..167..201S}, this makes this object the first extragalactic SySt with a resolved nebula.

\begin{figure}
  \resizebox{\hsize}{!}{\includegraphics{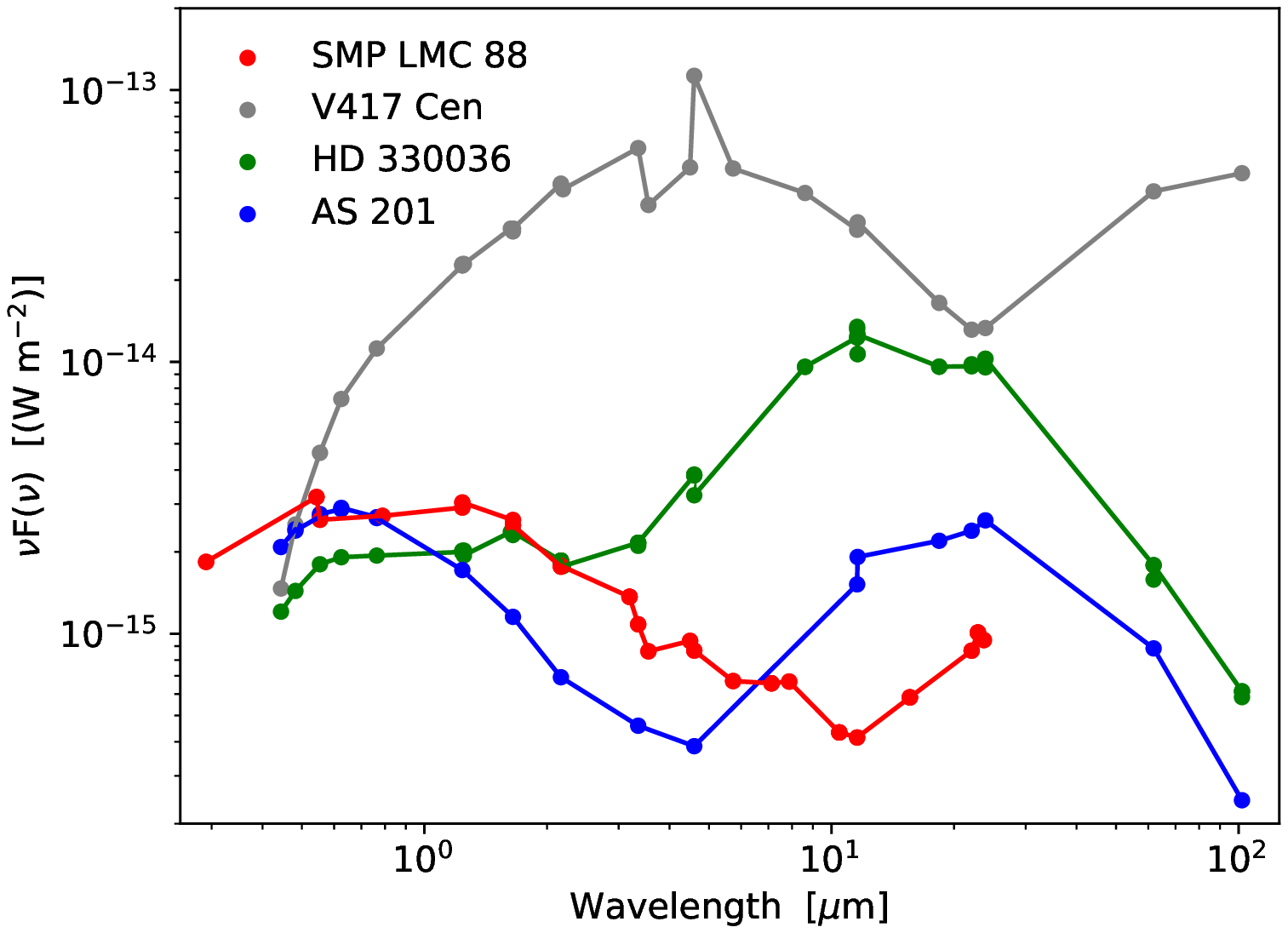}}
  \caption{The SED of SMP~LMC~88 compared with SEDs of the Galactic D'-type SySt. Flux of the Galactic SySt has been scaled to the distance of LMC \citep{2013Natur.495...76P}. The adopted distances of Galactic SySt are 5~kpc for V417~Cen \citep{1994A&A...285..241V}, 2.3~kpc for HD~330036 \citep{2005A&A...429..993P} and 4.3~kpc for AS~201 \citep{2005A&A...429..993P}. The observations are from the Two Micron All Sky Survey  \citep[2MASS;][]{2006AJ....131.1163S}, the Wide-field Infrared Survey Explorer \citep[WISE;][]{2010AJ....140.1868W}, AKARI/IRC mid-infrared all-sky survey \citep{2010A&A...514A...1I}, AKARI-LMC Point-source catalog \citep{2012AJ....144..179K}, Spitzer \citep{2008AJ....135..726H}, XMM-Newton Optical Monitor \citep{2012MNRAS.426..903P}, AAVSO Photometric All Sky Survey (APASS; \citealt{2012JAVSO..40..430H})  and OGLE \citep{2012AcA....62..247U}.   }
  \label{sed_fig}
\end{figure}

\begin{figure}
  \resizebox{\hsize}{!}{\includegraphics{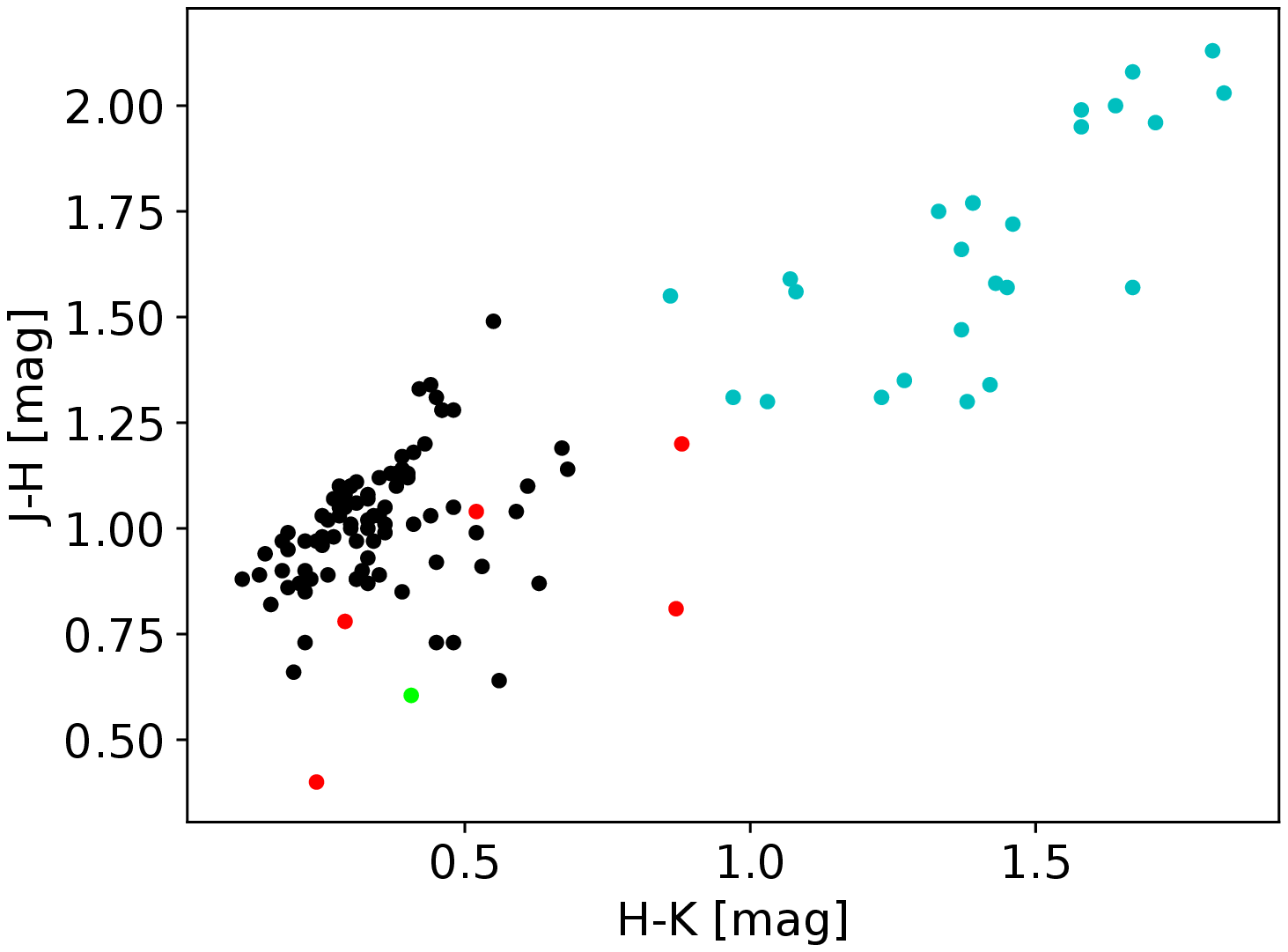}}
  \caption{IR colors of S-type (black points), D-type (blue points) and D'-type (red points) SySt from \citet{1992A&AS...93..383M} and 2MASS colors of SMP~LMC~88 (green point).}
  \label{2mass_fig}
\end{figure}

In the spectral energy distribution (SED) of SMP~LMC~88 there is an excess emission from dust in the IR  (Fig.~\ref{sed_fig}). The presence of dust means that the SySt type should be either a D- or D'-type SySt, no matter what is the source of dust. The IR colors of the system exclude a D-type classification of SMP~LMC~88 (Fig.~\ref{2mass_fig}), hence the system is a D'-type SySt.  The D'-type classification of SMP~LMC~88 is further supported by the comparison of its  SED with SEDs of other D'-type systems (Fig.~\ref{sed_fig}). Namely, the optical brightness is similar, and consistent with the presence of a giant. Moreover, the dust emission reaches its maximum longwards $\sim$10$\mu$m. This means that the dust is relatively cold ($<$400~K), which is in agreement with the classical definition of a D'-type SySt, where the D- and D'-type SySt were initially only distinguished by the significantly colder dust in the latter  \citep{1982ASSL...95...27A}. The D'-type systems with resolved optical nebulae are extremely rare, with only two such objects known in the Galaxy \citep[e.g.][and references therein]{{2008A&A...485..117S}}. Additionally, in such systems the nebula is thought to be a fossil AGB remnant, thus the SySt in SMP~LMC~88 LMC 88 can also be classified as a bona fide binary central star of a PN.

\begin{figure}
  \resizebox{\hsize}{!}{\includegraphics{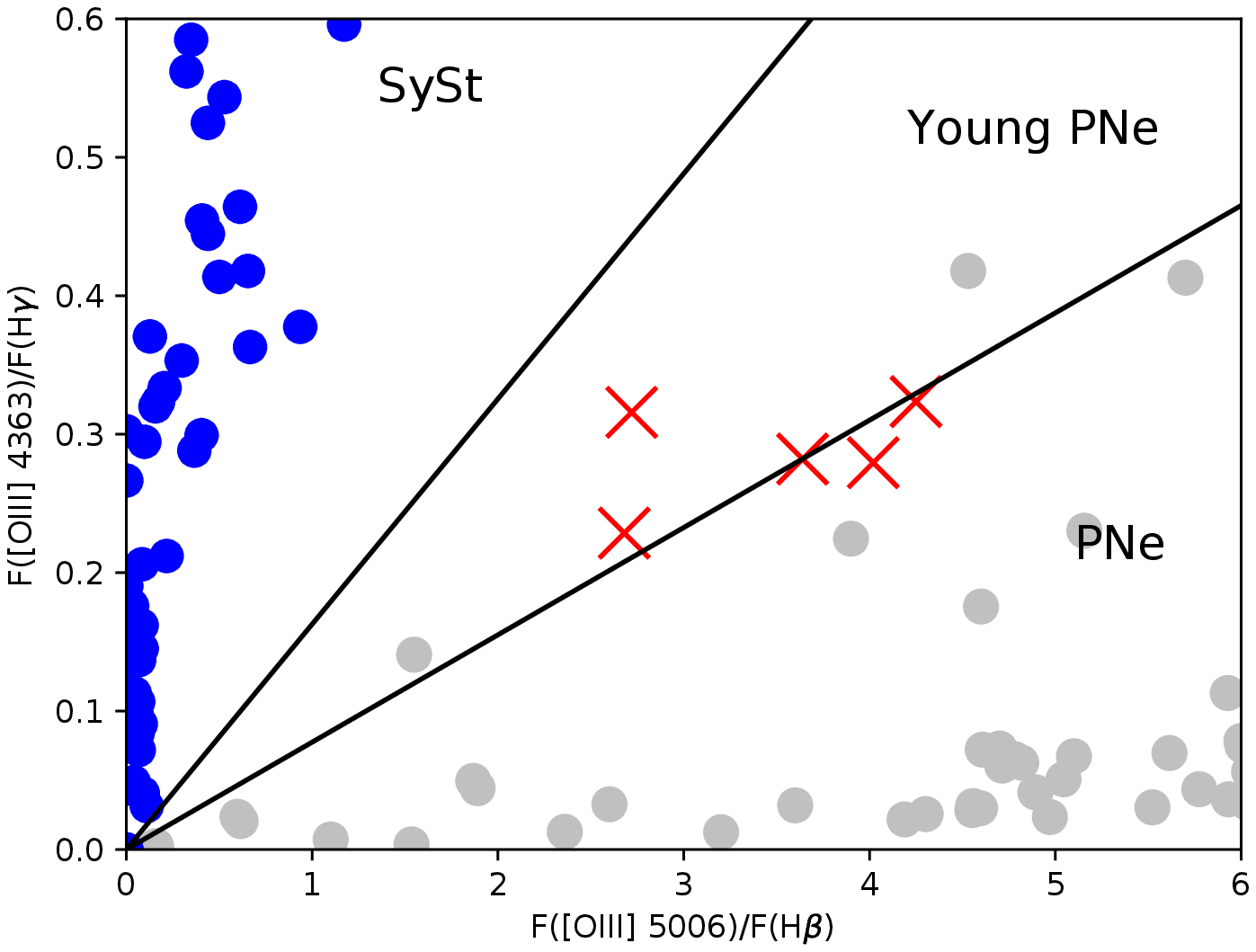}}
  \resizebox{\hsize}{!}{\includegraphics{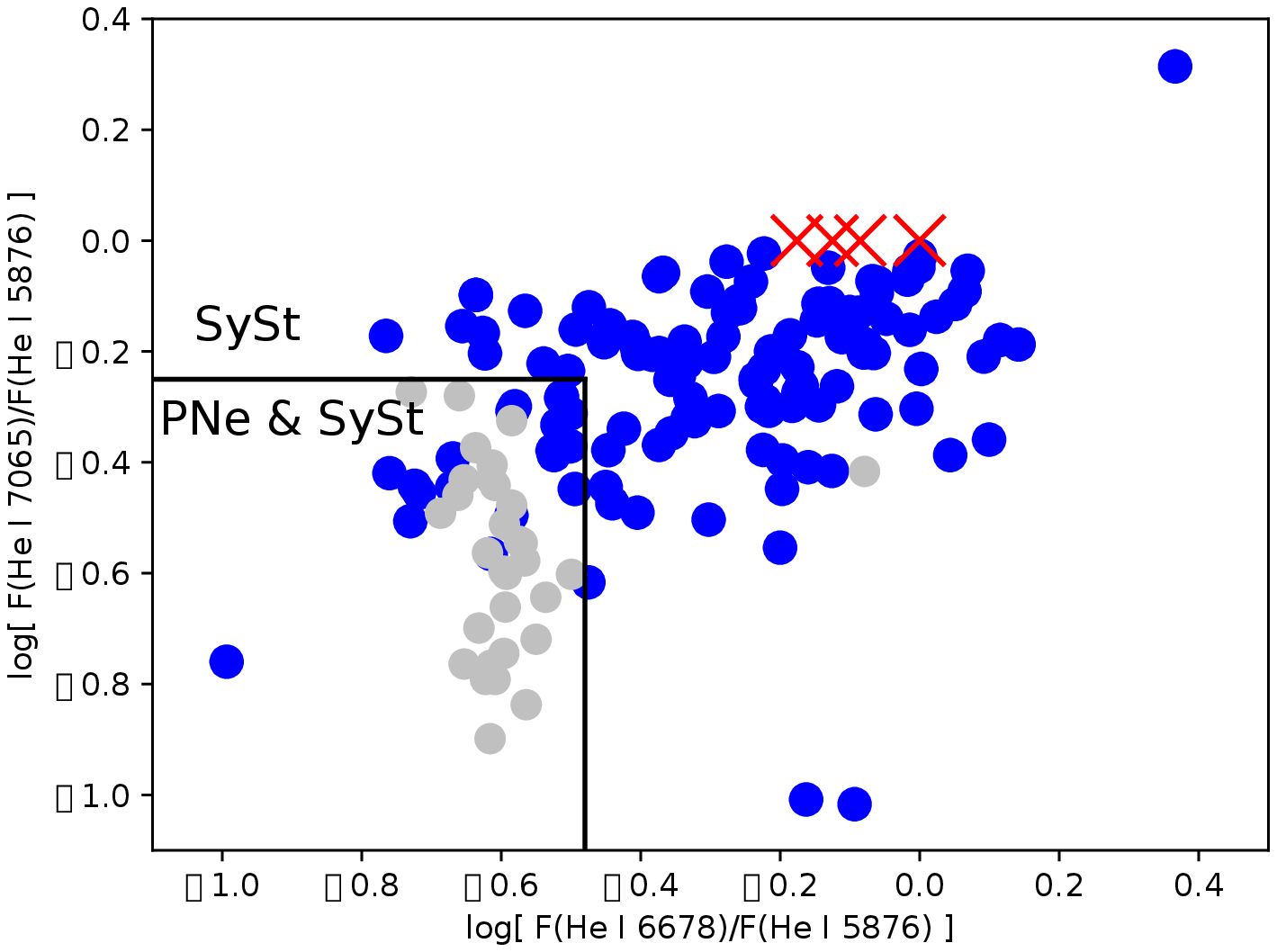}}

  \caption{The position of SMP~LMC~88 on the [\mbox{O\,{\sc iii}}] and \mbox{He\,{\sc i}} diagnostic diagrams with the observed line ratios from Table~\ref{table:lineratios} (red crosses). The Galactic SySt are marked with blue circles and the Galactic PNe are marked with gray circles. The data for Galactic systems are from \citep{2017A&A...606A.110I}.  } 
  \label{diagnostic_diagram}
\end{figure}

The SySt classification can be assessed in a diagnostic diagram with [\mbox{O\,{\sc iii}}] and Balmer lines flux ratios \hbox{\citep{1995PASP..107..462G}}. SMP~LMC~88 on this diagnostic diagram is placed between PNe and young PNe (Fig.~\ref{diagnostic_diagram}). This does not contradict the SySt classification since some SySt have been observed outside of the SySt region of this diagram \citep{2010IAUS..262..307B}. Since the \citet{1995PASP..107..462G} diagram is essentially a measure of density in the nebula, the position of SMP~LMC~88 may suggest that the [\mbox{O\,{\sc iii}}] emission is originating in a genuine PN rather than in the mass donor wind ionized by accreting WD.  Another available diagnostic diagram is a \mbox{He\,{\sc i}} diagram \citep{2017A&A...606A.110I}, which is not as likely to be affected by the PN ejecta, since the employed emission line form in a region is much closer to the central object that is dominated by the donor wind. In fact, the position of SMP~LMC~88 on the \mbox{He\,{\sc i}} diagram confirms its symbiotic nature (Fig.~\ref{diagnostic_diagram}). The different position of SMP~LMC~88 on [\mbox{O\,{\sc iii}}] and \mbox{He\,{\sc i}} diagrams is most probably due to the fact that the \mbox{He\,{\sc i}} emission lines form in the central region of the binary, while the [\mbox{O\,{\sc iii}}] emission lines mainly form in the resolved nebula.

\begin{figure*}
  \resizebox{\hsize}{!}{\includegraphics{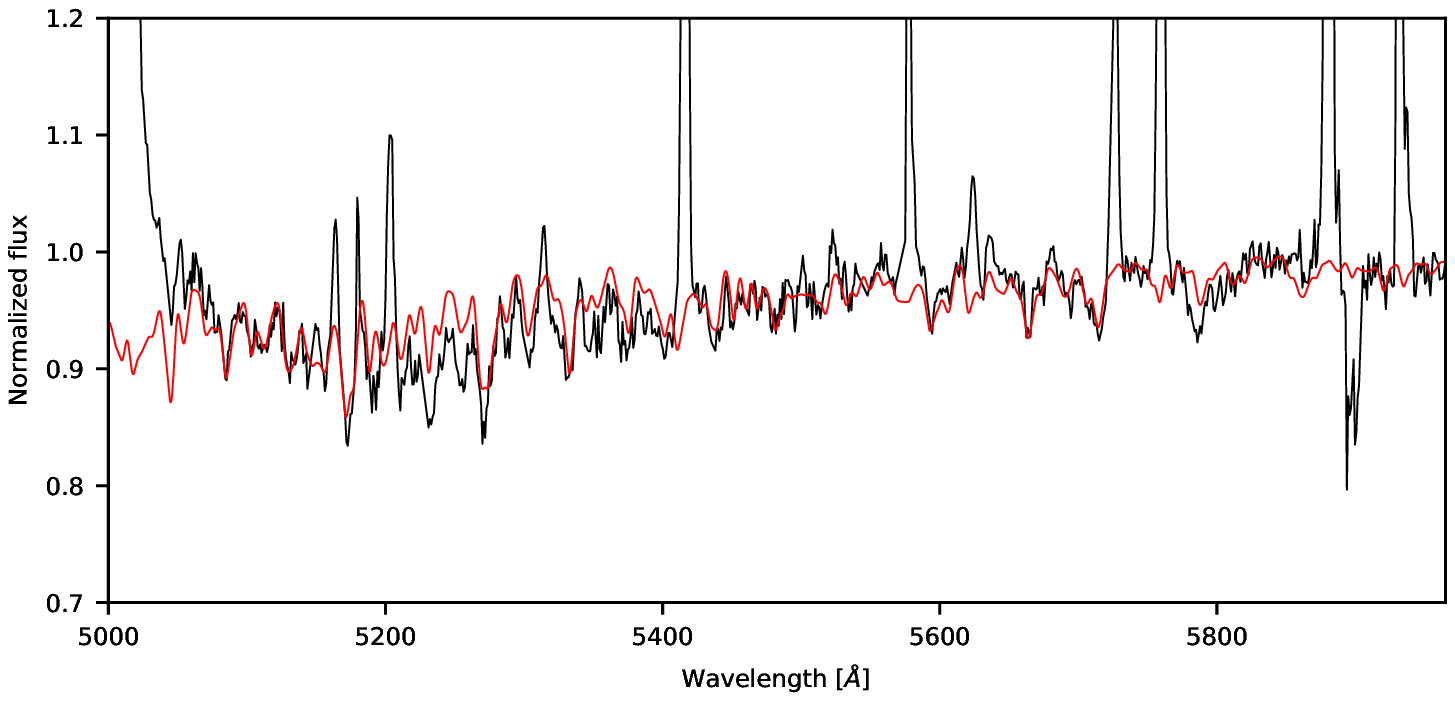}}
  \caption{Comparison of SALT 2017 spectrum (black line) and a synthetic spectrum of a giant with T$_{eff}$=5000~K. The spectra were normalized to the local continuum. The Na\,{\sc i}~D line is probably of interstellar origin. }
  \label{SMP88_synthetic}
\end{figure*}

\citet{1996A&AS..116...95L} noted a red continuum characteristic of a cold star with T$_{eff}$<4000--5000~K. In order to confirm presence of a cold star in SMP~LMC~88 we used a synthetic spectrum and compared it to our deepest spectrum in a spectral region with relatively low number of emission lines, as well as high number of expected absorption lines. The synthetic spectrum was calculated with mean LMC metallicity $\mathrm{[M/H]}=-0.5$ using SPECTRUM code \citep{1994AJ....107..742G} and \citet{2003IAUS..210P.A20C} stellar model. We assumed a $\log \mathrm{g}$=2.0 and T$_{eff}$=5000~K, typical for a giant in a D'-type SySt and consistent with \citet{1996A&AS..116...95L} observation. The comparison confirms presence of a cold star in the system (Fig.~\ref{SMP88_synthetic}). Due to non-negligible contribution of nebular spectrum as well as the hot component spectrum to the total continuum we were unable to precisely estimate the cold star T$_{eff}$. However, the comparison to synthetic spectrum, as well as presence of prominent G-band in our spectra, is enough to classify the star as a K-type star. This spectral type is later than in a typical D'-type SySt, but is similar to a yellow type SySt V417 Cen \citep[see e.g.][and references therein]{1999A&AS..137..473M}.

\subsection{Nebula properties}\label{neb_sec}


Due to the fact that SMP~LMC~88 was classified as a PN, the nebula properties were analysed extensively in the literature. The main problem with the nebular characteristics derived by different authors is due to that they used significantly different reddening values, ranging from c=0.58 \citep{2006ApJS..167..201S}, c=0.77 \citep{1991ApJS...75..407M} up to c=1.2 \citep{1996A&AS..116...95L} and c=1.31 \citep{2006AA...456..451L}. This is due to the fact that the reddening parameter c is measured using Balmer lines ratios and assuming Case~B. The Balmer line ratios vary significantly in SMP~LMC~88 (see Table~\ref{table:lineratios}) which means that at least at some epochs the emission is inconsistent with Case~B and the reddening values derived from Balmer lines are significantly overestimated. Here we have adopted the foreground Galactic extinction,  A$_V$=0.168 from the \citet[c=0.07;][]{2011ApJ...737..103S}, which represents the lower limit to the reddening towards SMP LMC 88.

We derived T$_e$ and n$_e$ using the reddening corrected [\mbox{N\,{\sc ii}}]~5755/6584 and [\mbox{S\,{\sc ii}}]~6731/6716 line ratios. Using the AAT spectrum we obtained T$_e$=13900~K and n$_e$=3900~cm$^{-3}$, while with the SALT spectrum we obtained   T$_e$=12400~K and n$_e$=3700~cm$^{-3}$. We estimated M$_i$ and $\varepsilon$ using the model of \citet{1994A&A...284..248B}. We assumed the nebular radius of 0.6~arcsec (\mbox{\citealt{1990ApJ...365..471J}}; \mbox{\citealt{2006ApJS..167..201S}}) and the distance to LMC 49.97~kpc \citep{2013Natur.495...76P}.  The flux calibration has been obtained by scaling the spectra to the mean OGLE~$V$ magnitude, which resulted in F(H$\beta$)=2.6$\times10^{-14}$ in the AAT spectrum and F(H$\beta$)=1.7$\times10^{-14}$~erg~cm$^{-2}$~s$^{-1}$ in the 2015~SALT spectrum. Using a reddening corrected H$\beta$ flux, T$_e$ and n$_e$ from the AAT spectrum we derive M$_i$=0.03~M$_\odot$ and $\varepsilon$=0.17, which indicates a total mass of the nebula M$_{neb}$=0.18~M$_\odot$. Using the 2015~SALT spectrum we derive M$_i$=0.02~M$_\odot$ and $\varepsilon$=0.11, which indicates M$_{neb}$=0.17~M$_\odot$. There is a big discrepancy between the ionized mass derived in this work and published in the literature (M$_i$=0.575~M$_\odot$ of \citealt{1991ApJ...367..115D} and M$_i$=0.343~M$_\odot$ of \citealt{1994A&A...284..248B}) which can be mainly attributed to difference in adopted reddening values. Namely, if we assumed the same reddening as \citet{1991ApJ...367..115D} and \citet{1994A&A...284..248B}, we would get M$_i$=0.16~M$_\odot$. Moreover, the emission lines are variable, which would naturally cause differences in the derived nebula parameters when observations from different epochs are used.

Two components of the forbidden emission lines of the nebula are clearly separated on the HST/STIS spectrum (Fig.~\ref{HST_lines}). The [\mbox{O\,{\sc iii}}] lines have an expansion velocity  v$_{exp}\simeq$33~km/s, while the [\mbox{N\,{\sc ii}}] lines have v$_{exp}\simeq$60~km/s. This is consistent with the fact that \citet{2006ApJS..167..201S} observed larger and more pronounced ring of emission in [\mbox{N\,{\sc ii}}] than in [\mbox{O\,{\sc iii}}]. The wavelength calibration of the HST/STIS spectrum seems of low quality, however the difference in expansion velocities is confirmed by differences in the widths of lines on the AAT spectrum. The [\mbox{O\,{\sc iii}}] expansion velocity if roughly consistent with v$_{exp}=$24.7~km/s of \citet{1988ApJ...327..639D} estimated based on width of the line profile in their spectrum. When we assume a lower and upper limit of the size of the nebula to be 0.45 and 0.61~arcsec respectively, a distance of 49.97~kpc \citep{2013Natur.495...76P}, and we take a lower and upper limit on the expansion velocity of 33 and 60~km/s respectively, the estimated dynamical age of the nebula is 1800--4400~yrs.

The parameters of the nebula of SMP~LMC~88 are similar to the nebula of AS~201 (Table~\ref{table:comparison}),  one of the only two D'-type SySt with a resolved nebula known thus far. The similarity of the two nebulae confirms the D'-type classification of SMP~LMC~88 because nebulae of symbiotic Miras typically have significantly lower ionized masses, with only one exception of Hen~2-104 \citep{2008A&A...485..117S}.

\begin{table}
\caption{Comparison of the SMP~LMC~88 nebula (see text) to the parameters of the nebula in AS~201 derived by \citet{1991A&A...243..469S}.}         
\label{table:comparison}     
\centering                       
\begin{tabular}{ccc}       
\hline\hline   
      & SMP~LMC~88 &    AS~201\\
\hline
  M$_i$ & 0.02--0.03~M$_\odot$  & 0.08~M$_\odot$ \\
  Size & 0.15$\times$0.11~pc  & 0.1~pc \\
  Age & 1800--4400~yrs & 4000~yrs\\
\hline                    
\end{tabular}
\end{table}

\subsection{Variability}\label{var_sec}

SMP~LMC~88 showed a sinusoidal variability in the OGLE data  (Fig.~\ref{SMP88_ogle}). The amplitude of variability is $\sim$0.1~mag in the OGLE $I$ filter. The variability might be induced by orbital motion. If this is the case, the orbital period of SMP~LMC~88 would be $\sim$4900d, as suggested by separation of the two maxima. This is supported by the fact that similar sinusoidal variability related to the orbital motion with a period of order of a few thousands of days and with an amplitude of order of 0.1~mag is present in the Galactic D'-type SySt HD~330036 and Hen~3-1591 \citep{2013AcA....63..405G}.  Another possibility would be that the star exhibited an active phase characteristic for a SySt, for which the amplitude of variability is smaller at redder wavelengths \citep[see e.g.][]{2016MNRAS.462.2695I}. This would be consistent with the small amplitude of SMP~LMC~88 variability in the OGLE $I$ filter.

The forbidden to allowed emission lines flux ratios are clearly variable (Fig.~\ref{SMP88_raman}). In particular the [\mbox{O\,{\sc iii}]\,5007}/H$\beta$ ratio reported by different authors varies between $\sim$3 and $\sim$4. Similarly the  H$\alpha$/H$\beta$  ratio varies between $\sim$4.5 and $\sim$9. It is not clear what is the nature of this variability or whether the variability is correlated with the photometric observations. However, the variability is clearly not related to changes in the absorbing column density, since line ratios variability are not correlated with wavelength.

Both the photometric and spectroscopic variability could not be reproduced by a single nucleus of PN, which supports the SySt classification of SMP~LMC~88.

\section{Summary}\label{sumarysec}

In this work we identified Raman scattered \mbox{O\,{\sc vi}} lines in SMP~LMC~88, which shows that the system is a SySt.  The SySt classification is supported by the  \mbox{He\,{\sc i}} diagnostic diagram \citep{2017A&A...606A.110I}. The IR colors and stellar features in the continuum allowed us to classified the cold component as a K-type giant, which means that the system is as a yellow D'-type SySt. Such systems are rare and SMP~LMC~88 is the first such system discovered outside of the Galaxy. Moreover, SMP~LMC~88 has an optically resolved nebula, which is the first such nebula observed in an extragalactic SySt. The only resolved feature observed in extragalactic SySt thus far is the giant jet in  Sanduleak's star \citep{2011ApJ...743L...8A}. Moreover, the resolved nebulae in the yellow D'-type SySt are extremely rare in general, with only two such objects known in the Galaxy \citep[see e.g.][and references therein]{2003ASPC..303..393C,2008A&A...485..117S}. 

The fact that the nebula around a yellow D'-type SySt, opposite to the nebulae around D-type SySt, is an actual PN (i.e. a fossil AGB remnant) makes the central binary of SMP~LMC~88 a central binary of a PN. Due to the fact that central stars of PNe are discovered mainly using photometric monitoring, the known sample of central binary stars of PNe consists mainly of binaries with short orbital periods \citep{2009A&A...496..813M}. On the other hand SySt are binaries with some of the longest orbital periods, which makes SMP~LMC~88 an important addition to the known population of binary central stars of PNe as the importance of long-period central stars of PNe remains poorly understood \citep[see][and references therein]{2014A&A...563L..10V,2017A&A...600L...9J,2018MNRAS.473.2275M}.

We showed that there is evidence of variability in the photometric observations, as well as in emission line fluxes. The photometric variability is similar to orbital variability of the Galactic D'-type SySt. If the observed variability is in fact due to orbital motion the orbital period would be $\sim$4900d. More observations are needed to determine the nature of the spectroscopic variability.

\section*{Acknowledgements}
We are grateful to Romano Corradi for an inspiring discussion. This study has been supported in part by the Polish MNiSW grant 0136/DIA/2014/43, and NCN grant DEC-2013/10/M/ST9/00086. Polish participation in SALT is funded by grant No. MNiSW DIR/WK/2016/07. The OGLE project has received funding from the National Science Centre, Poland, grant MAESTRO 2014/14/A/ST9/00121 to A.~Udalski. BM acknowledges support from the National Research Foundation (NRF) of South Africa. This research has made use of the VizieR catalogue access tool, CDS, Strasbourg, France. The original description of the VizieR service was published in A\&AS 143, 23.




\bibliographystyle{mnras}
\bibliography{references} 

\begin{thebibliography}{}
\makeatletter
\relax
\def\mn@urlcharsother{\let\do\@makeother \do\$\do\&\do\#\do\^\do\_\do\%\do\~}
\def\mn@doi{\begingroup\mn@urlcharsother \@ifnextchar [ {\mn@doi@}
  {\mn@doi@[]}}
\def\mn@doi@[#1]#2{\def\@tempa{#1}\ifx\@tempa\@empty \href
  {http://dx.doi.org/#2} {doi:#2}\else \href {http://dx.doi.org/#2} {#1}\fi
  \endgroup}
\def\mn@eprint#1#2{\mn@eprint@#1:#2::\@nil}
\def\mn@eprint@arXiv#1{\href {http://arxiv.org/abs/#1} {{\tt arXiv:#1}}}
\def\mn@eprint@dblp#1{\href {http://dblp.uni-trier.de/rec/bibtex/#1.xml}
  {dblp:#1}}
\def\mn@eprint@#1:#2:#3:#4\@nil{\def\@tempa {#1}\def\@tempb {#2}\def\@tempc
  {#3}\ifx \@tempc \@empty \let \@tempc \@tempb \let \@tempb \@tempa \fi \ifx
  \@tempb \@empty \def\@tempb {arXiv}\fi \@ifundefined
  {mn@eprint@\@tempb}{\@tempb:\@tempc}{\expandafter \expandafter \csname
  mn@eprint@\@tempb\endcsname \expandafter{\@tempc}}}

\bibitem[\protect\citeauthoryear{{Allen}}{{Allen}}{1982}]{1982ASSL...95...27A}
{Allen} D.~A.,  1982, in {Friedjung} M.,  {Viotti} R.,  eds,  Astrophysics and
  Space Science Library Vol. 95, IAU Colloq. 70: The Nature of Symbiotic Stars.
  pp 27--42, \mn@doi{10.1007/978-94-009-7834-8_4}

\bibitem[\protect\citeauthoryear{{Angeloni}, {Di Mille}, {Bland-Hawthorn}  \&
  {Osip}}{{Angeloni} et~al.}{2011}]{2011ApJ...743L...8A}
{Angeloni} R.,  {Di Mille} F.,  {Bland-Hawthorn} J.,   {Osip} D.~J.,  2011,
  \mn@doi [\apjl] {10.1088/2041-8205/743/1/L8}, \href
  {http://adsabs.harvard.edu/abs/2011ApJ...743L...8A} {743, L8}

\bibitem[\protect\citeauthoryear{{Baella}}{{Baella}}{2010}]{2010IAUS..262..307B}
{Baella} N.~O.,  2010, in {Bruzual} G.~R.,  {Charlot} S.,  eds,  IAU Symposium
  Vol. 262, Stellar Populations - Planning for the Next Decade. pp 307--308,
  \mn@doi{10.1017/S1743921310002978}

\bibitem[\protect\citeauthoryear{{Boffi} \& {Stanghellini}}{{Boffi} \&
  {Stanghellini}}{1994}]{1994A&A...284..248B}
{Boffi} F.~R.,  {Stanghellini} L.,  1994, \aap, \href
  {http://adsabs.harvard.edu/abs/1994A%26A...284..248B} {284, 248}

\bibitem[\protect\citeauthoryear{{Burgh}, {Nordsieck}, {Kobulnicky},
  {Williams}, {O'Donoghue}, {Smith}  \& {Percival}}{{Burgh}
  et~al.}{2003}]{2003SPIE.4841.1463B}
{Burgh} E.~B.,  {Nordsieck} K.~H.,  {Kobulnicky} H.~A.,  {Williams} T.~B.,
  {O'Donoghue} D.,  {Smith} M.~P.,   {Percival} J.~W.,  2003, in {Iye} M.,
  {Moorwood} A.~F.~M.,  eds,  \procspie Vol. 4841, Instrument Design and
  Performance for Optical/Infrared Ground-based Telescopes. pp 1463--1471,
  \mn@doi{10.1117/12.460312}

\bibitem[\protect\citeauthoryear{{Castelli} \& {Kurucz}}{{Castelli} \&
  {Kurucz}}{2003}]{2003IAUS..210P.A20C}
{Castelli} F.,  {Kurucz} R.~L.,  2003, in {Piskunov} N.,  {Weiss} W.~W.,
  {Gray} D.~F.,  eds,  IAU Symposium Vol. 210, Modelling of Stellar
  Atmospheres. p.~A20

\bibitem[\protect\citeauthoryear{{Corradi}}{{Corradi}}{2003}]{2003ASPC..303..393C}
{Corradi} R.~L.~M.,  2003, in {Corradi} R.~L.~M.,  {Mikolajewska} J.,
  {Mahoney} T.~J.,  eds,  Astronomical Society of the Pacific Conference Series
  Vol. 303, Symbiotic Stars Probing Stellar Evolution. p.~393

\bibitem[\protect\citeauthoryear{{Corradi} \& {Schwarz}}{{Corradi} \&
  {Schwarz}}{1997}]{1997ppsb.conf..147C}
{Corradi} R.,  {Schwarz} H.~E.,  1997, in {Miko{\l}ajewska} J.,  ed., Physical
  Processes in Symbiotic Binaries and Related Systems. p.~147

\bibitem[\protect\citeauthoryear{{Corradi}, {Brandi}, {Ferrer}  \&
  {Schwarz}}{{Corradi} et~al.}{1999a}]{1999A&A...343..841C}
{Corradi} R.~L.~M.,  {Brandi} E.,  {Ferrer} O.~E.,   {Schwarz} H.~E.,  1999a,
  \aap, \href {http://adsabs.harvard.edu/abs/1999A%26A...343..841C} {343, 841}

\bibitem[\protect\citeauthoryear{{Corradi}, {Ferrer}, {Schwarz}, {Brandi}  \&
  {Garc{\'{\i}}a}}{{Corradi} et~al.}{1999b}]{1999A&A...348..978C}
{Corradi} R.~L.~M.,  {Ferrer} O.~E.,  {Schwarz} H.~E.,  {Brandi} E.,
  {Garc{\'{\i}}a} L.,  1999b, \aap, \href
  {http://adsabs.harvard.edu/abs/1999A%26A...348..978C} {348, 978}

\bibitem[\protect\citeauthoryear{{Crawford} et~al.,}{{Crawford}
  et~al.}{2010}]{2010SPIE.7737E..25C}
{Crawford} S.~M.,  et~al., 2010, in Observatory Operations: Strategies,
  Processes, and Systems III. p. 773725, \mn@doi{10.1117/12.857000}

\bibitem[\protect\citeauthoryear{{Dopita} \& {Meatheringham}}{{Dopita} \&
  {Meatheringham}}{1991}]{1991ApJ...367..115D}
{Dopita} M.~A.,  {Meatheringham} S.~J.,  1991, \mn@doi [\apj] {10.1086/169607},
  \href {http://adsabs.harvard.edu/abs/1991ApJ...367..115D} {367, 115}

\bibitem[\protect\citeauthoryear{{Dopita}, {Meatheringham}, {Webster}  \&
  {Ford}}{{Dopita} et~al.}{1988}]{1988ApJ...327..639D}
{Dopita} M.~A.,  {Meatheringham} S.~J.,  {Webster} B.~L.,   {Ford} H.~C.,
  1988, \mn@doi [\apj] {10.1086/166221}, \href
  {http://adsabs.harvard.edu/abs/1988ApJ...327..639D} {327, 639}

\bibitem[\protect\citeauthoryear{{Gray} \& {Corbally}}{{Gray} \&
  {Corbally}}{1994}]{1994AJ....107..742G}
{Gray} R.~O.,  {Corbally} C.~J.,  1994, \mn@doi [\aj] {10.1086/116893}, \href
  {http://adsabs.harvard.edu/abs/1994AJ....107..742G} {107, 742}

\bibitem[\protect\citeauthoryear{{Gromadzki}, {Miko{\l}ajewska}  \&
  {Soszy{\'n}ski}}{{Gromadzki} et~al.}{2013}]{2013AcA....63..405G}
{Gromadzki} M.,  {Miko{\l}ajewska} J.,   {Soszy{\'n}ski} I.,  2013, \actaa,
  \href {http://adsabs.harvard.edu/abs/2013AcA....63..405G} {63, 405}

\bibitem[\protect\citeauthoryear{{Gutierrez-Moreno}, {Moreno}  \&
  {Cortes}}{{Gutierrez-Moreno} et~al.}{1995}]{1995PASP..107..462G}
{Gutierrez-Moreno} A.,  {Moreno} H.,   {Cortes} G.,  1995, \mn@doi [\pasp]
  {10.1086/133575}, \href {http://adsabs.harvard.edu/abs/1995PASP..107..462G}
  {107, 462}

\bibitem[\protect\citeauthoryear{{Henden}, {Levine}, {Terrell}, {Smith}  \&
  {Welch}}{{Henden} et~al.}{2012}]{2012JAVSO..40..430H}
{Henden} A.~A.,  {Levine} S.~E.,  {Terrell} D.,  {Smith} T.~C.,   {Welch} D.,
  2012, Journal of the American Association of Variable Star Observers
  (JAAVSO), \href {http://adsabs.harvard.edu/abs/2012JAVSO..40..430H} {40, 430}

\bibitem[\protect\citeauthoryear{{Hora} et~al.,}{{Hora}
  et~al.}{2008}]{2008AJ....135..726H}
{Hora} J.~L.,  et~al., 2008, \mn@doi [\aj] {10.1088/0004-6256/135/2/726}, \href
  {http://cdsads.u-strasbg.fr/abs/2008AJ....135..726H} {135, 726}

\bibitem[\protect\citeauthoryear{{I{\l}kiewicz} \&
  {Miko{\l}ajewska}}{{I{\l}kiewicz} \&
  {Miko{\l}ajewska}}{2017}]{2017A&A...606A.110I}
{I{\l}kiewicz} K.,  {Miko{\l}ajewska} J.,  2017, \mn@doi [\aap]
  {10.1051/0004-6361/201731497}, \href
  {http://adsabs.harvard.edu/abs/2017A%26A...606A.110I} {606, A110}

\bibitem[\protect\citeauthoryear{{I{\l}kiewicz}, {Miko{\l}ajewska}, {Stoyanov},
  {Manousakis}  \& {Miszalski}}{{I{\l}kiewicz}
  et~al.}{2016}]{2016MNRAS.462.2695I}
{I{\l}kiewicz} K.,  {Miko{\l}ajewska} J.,  {Stoyanov} K.,  {Manousakis} A.,
  {Miszalski} B.,  2016, \mn@doi [\mnras] {10.1093/mnras/stw1837}, \href
  {http://adsabs.harvard.edu/abs/2016MNRAS.462.2695I} {462, 2695}

\bibitem[\protect\citeauthoryear{{Ishihara} et~al.,}{{Ishihara}
  et~al.}{2010}]{2010A&A...514A...1I}
{Ishihara} D.,  et~al., 2010, \mn@doi [\aap] {10.1051/0004-6361/200913811},
  \href {http://cdsads.u-strasbg.fr/abs/2010A%26A...514A...1I} {514, A1}

\bibitem[\protect\citeauthoryear{{Jacoby}, {Ciardullo}  \& {Walker}}{{Jacoby}
  et~al.}{1990}]{1990ApJ...365..471J}
{Jacoby} G.~H.,  {Ciardullo} R.,   {Walker} A.~R.,  1990, \mn@doi [\apj]
  {10.1086/169501}, \href {http://adsabs.harvard.edu/abs/1990ApJ...365..471J}
  {365, 471}

\bibitem[\protect\citeauthoryear{{Jones}, {Van Winckel}, {Aller}, {Exter}  \&
  {De Marco}}{{Jones} et~al.}{2017}]{2017A&A...600L...9J}
{Jones} D.,  {Van Winckel} H.,  {Aller} A.,  {Exter} K.,   {De Marco} O.,
  2017, \mn@doi [\aap] {10.1051/0004-6361/201730700}, \href
  {http://adsabs.harvard.edu/abs/2017A%26A...600L...9J} {600, L9}

\bibitem[\protect\citeauthoryear{{Kang}, {Lee}  \& {Lee}}{{Kang}
  et~al.}{2009}]{2009ApJ...695..542K}
{Kang} E.-H.,  {Lee} B.-C.,   {Lee} H.-W.,  2009, \mn@doi [\apj]
  {10.1088/0004-637X/695/1/542}, \href
  {http://adsabs.harvard.edu/abs/2009ApJ...695..542K} {695, 542}

\bibitem[\protect\citeauthoryear{{Kato} et~al.,}{{Kato}
  et~al.}{2012}]{2012AJ....144..179K}
{Kato} D.,  et~al., 2012, \mn@doi [\aj] {10.1088/0004-6256/144/6/179}, \href
  {http://cdsads.u-strasbg.fr/abs/2012AJ....144..179K} {144, 179}

\bibitem[\protect\citeauthoryear{{Kimble} et~al.,}{{Kimble}
  et~al.}{1998}]{1998ApJ...492L..83K}
{Kimble} R.~A.,  et~al., 1998, \mn@doi [\apjl] {10.1086/311102}, \href
  {http://adsabs.harvard.edu/abs/1998ApJ...492L..83K} {492, L83}

\bibitem[\protect\citeauthoryear{{Kobulnicky}, {Nordsieck}, {Burgh}, {Smith},
  {Percival}, {Williams}  \& {O'Donoghue}}{{Kobulnicky}
  et~al.}{2003}]{2003SPIE.4841.1634K}
{Kobulnicky} H.~A.,  {Nordsieck} K.~H.,  {Burgh} E.~B.,  {Smith} M.~P.,
  {Percival} J.~W.,  {Williams} T.~B.,   {O'Donoghue} D.,  2003, in {Iye} M.,
  {Moorwood} A.~F.~M.,  eds,  \procspie Vol. 4841, Instrument Design and
  Performance for Optical/Infrared Ground-based Telescopes. pp 1634--1644,
  \mn@doi{10.1117/12.460315}

\bibitem[\protect\citeauthoryear{{Kontizas}, {Morgan}, {Kontizas}  \&
  {Dapergolas}}{{Kontizas} et~al.}{1996}]{1996A&A...307..359K}
{Kontizas} M.,  {Morgan} D.~H.,  {Kontizas} E.,   {Dapergolas} A.,  1996, \aap,
  \href {http://adsabs.harvard.edu/abs/1996A%26A...307..359K} {307, 359}

\bibitem[\protect\citeauthoryear{{Koz{\l}owski} et~al.,}{{Koz{\l}owski}
  et~al.}{2013}]{2013ApJ...775...92K}
{Koz{\l}owski} S.,  et~al., 2013, \mn@doi [\apj] {10.1088/0004-637X/775/2/92},
  \href {http://adsabs.harvard.edu/abs/2013ApJ...775...92K} {775, 92}

\bibitem[\protect\citeauthoryear{{Kwok}}{{Kwok}}{2003}]{2003ASPC..303..428K}
{Kwok} S.,  2003, in {Corradi} R.~L.~M.,  {Mikolajewska} J.,   {Mahoney} T.~J.,
   eds,  Astronomical Society of the Pacific Conference Series Vol. 303,
  Symbiotic Stars Probing Stellar Evolution. p.~428

\bibitem[\protect\citeauthoryear{{Lee}, {Kang}  \& {Byun}}{{Lee}
  et~al.}{2001}]{2001ApJ...551L.121L}
{Lee} H.-W.,  {Kang} Y.-W.,   {Byun} Y.-I.,  2001, \mn@doi [\apjl]
  {10.1086/319830}, \href {http://adsabs.harvard.edu/abs/2001ApJ...551L.121L}
  {551, L121}

\bibitem[\protect\citeauthoryear{{Lee}, {Jung}, {Song}  \& {Ahn}}{{Lee}
  et~al.}{2006}]{2006ApJ...636.1045L}
{Lee} H.-W.,  {Jung} Y.-C.,  {Song} I.-O.,   {Ahn} S.-H.,  2006, \mn@doi [\apj]
  {10.1086/498143}, \href {http://adsabs.harvard.edu/abs/2006ApJ...636.1045L}
  {636, 1045}

\bibitem[\protect\citeauthoryear{{Leisy} \& {Dennefeld}}{{Leisy} \&
  {Dennefeld}}{1996}]{1996A&AS..116...95L}
{Leisy} P.,  {Dennefeld} M.,  1996, \aaps, \href
  {http://adsabs.harvard.edu/abs/1996A%26AS..116...95L} {116, 95}

\bibitem[\protect\citeauthoryear{{Leisy} \& {Dennefeld}}{{Leisy} \&
  {Dennefeld}}{2006}]{2006AA...456..451L}
{Leisy} P.,  {Dennefeld} M.,  2006, \mn@doi [\aap]
  {10.1051/0004-6361:20053063}, \href
  {http://cdsads.u-strasbg.fr/abs/2006A%26A...456..451L} {456, 451}

\bibitem[\protect\citeauthoryear{{McCollum}, {Bruhweiler}, {Wahlgren},
  {Eriksson}  \& {Verner}}{{McCollum} et~al.}{2008}]{2008ApJ...682.1087M}
{McCollum} B.,  {Bruhweiler} F.~C.,  {Wahlgren} G.~M.,  {Eriksson} M.,
  {Verner} E.,  2008, \mn@doi [\apj] {10.1086/589137}, \href
  {http://adsabs.harvard.edu/abs/2008ApJ...682.1087M} {682, 1087}

\bibitem[\protect\citeauthoryear{{Meatheringham} \& {Dopita}}{{Meatheringham}
  \& {Dopita}}{1991}]{1991ApJS...75..407M}
{Meatheringham} S.~J.,  {Dopita} M.~A.,  1991, \mn@doi [\apjs]
  {10.1086/191536}, \href {http://adsabs.harvard.edu/abs/1991ApJS...75..407M}
  {75, 407}

\bibitem[\protect\citeauthoryear{{Miko{\l}ajewska}}{{Miko{\l}ajewska}}{2012}]{2012BaltA..21....5M}
{Miko{\l}ajewska} J.,  2012, Baltic Astronomy, \href
  {http://adsabs.harvard.edu/abs/2012BaltA..21....5M} {21, 5}

\bibitem[\protect\citeauthoryear{{Miszalski}, {Acker}, {Moffat}, {Parker}  \&
  {Udalski}}{{Miszalski} et~al.}{2009}]{2009A&A...496..813M}
{Miszalski} B.,  {Acker} A.,  {Moffat} A.~F.~J.,  {Parker} Q.~A.,   {Udalski}
  A.,  2009, \mn@doi [\aap] {10.1051/0004-6361/200811380}, \href
  {http://adsabs.harvard.edu/abs/2009A%26A...496..813M} {496, 813}

\bibitem[\protect\citeauthoryear{{Miszalski}, {Boffin}, {Frew}, {Acker},
  {K{\"o}ppen}, {Moffat}  \& {Parker}}{{Miszalski}
  et~al.}{2012}]{2012MNRAS.419...39M}
{Miszalski} B.,  {Boffin} H.~M.~J.,  {Frew} D.~J.,  {Acker} A.,  {K{\"o}ppen}
  J.,  {Moffat} A.~F.~J.,   {Parker} Q.~A.,  2012, \mn@doi [\mnras]
  {10.1111/j.1365-2966.2011.19667.x}, \href
  {http://adsabs.harvard.edu/abs/2012MNRAS.419...39M} {419, 39}

\bibitem[\protect\citeauthoryear{{Miszalski}, {Manick}, {Miko{\l}ajewska},
  {I{\l}kiewicz}, {Kamath}  \& {Van Winckel}}{{Miszalski}
  et~al.}{2018}]{2018MNRAS.473.2275M}
{Miszalski} B.,  {Manick} R.,  {Miko{\l}ajewska} J.,  {I{\l}kiewicz} K.,
  {Kamath} D.,   {Van Winckel} H.,  2018, \mn@doi [\mnras]
  {10.1093/mnras/stx2501}, \href
  {http://adsabs.harvard.edu/abs/2018MNRAS.473.2275M} {473, 2275}

\bibitem[\protect\citeauthoryear{{Morgan}}{{Morgan}}{1984}]{1984MNRAS.208..633M}
{Morgan} D.~H.,  1984, \mn@doi [\mnras] {10.1093/mnras/208.3.633}, \href
  {http://adsabs.harvard.edu/abs/1984MNRAS.208..633M} {208, 633}

\bibitem[\protect\citeauthoryear{{Munari} \& {Patat}}{{Munari} \&
  {Patat}}{1993}]{1993A&A...277..195M}
{Munari} U.,  {Patat} F.,  1993, \aap, \href
  {http://adsabs.harvard.edu/abs/1993A%26A...277..195M} {277, 195}

\bibitem[\protect\citeauthoryear{{Munari}, {Yudin}, {Taranova}, {Massone},
  {Marang}, {Roberts}, {Winkler}  \& {Whitelock}}{{Munari}
  et~al.}{1992}]{1992A&AS...93..383M}
{Munari} U.,  {Yudin} B.~F.,  {Taranova} O.~G.,  {Massone} G.,  {Marang} F.,
  {Roberts} G.,  {Winkler} H.,   {Whitelock} P.~A.,  1992, \aaps, \href
  {http://adsabs.harvard.edu/abs/1992A%26AS...93..383M} {93, 383}

\bibitem[\protect\citeauthoryear{{M{\"u}rset} \& {Schmid}}{{M{\"u}rset} \&
  {Schmid}}{1999}]{1999A&AS..137..473M}
{M{\"u}rset} U.,  {Schmid} H.~M.,  1999, \mn@doi [\aaps] {10.1051/aas:1999105},
  \href {http://adsabs.harvard.edu/abs/1999A%26AS..137..473M} {137, 473}

\bibitem[\protect\citeauthoryear{{Nussbaumer}, {Schmid}  \&
  {Vogel}}{{Nussbaumer} et~al.}{1989}]{1989A&A...211L..27N}
{Nussbaumer} H.,  {Schmid} H.~M.,   {Vogel} M.,  1989, \aap, \href
  {http://adsabs.harvard.edu/abs/1989A%26A...211L..27N} {211, L27}

\bibitem[\protect\citeauthoryear{{O'Donoghue} et~al.,}{{O'Donoghue}
  et~al.}{2006}]{2006MNRAS.372..151O}
{O'Donoghue} D.,  et~al., 2006, \mn@doi [\mnras]
  {10.1111/j.1365-2966.2006.10834.x}, \href
  {http://adsabs.harvard.edu/abs/2006MNRAS.372..151O} {372, 151}

\bibitem[\protect\citeauthoryear{{Page} et~al.,}{{Page}
  et~al.}{2012}]{2012MNRAS.426..903P}
{Page} M.~J.,  et~al., 2012, \mn@doi [\mnras]
  {10.1111/j.1365-2966.2012.21706.x}, \href
  {http://cdsads.u-strasbg.fr/abs/2012MNRAS.426..903P} {426, 903}

\bibitem[\protect\citeauthoryear{{Pequignot}, {Baluteau}, {Morisset}  \&
  {Boisson}}{{Pequignot} et~al.}{1997}]{1997A&A...323..217P}
{Pequignot} D.,  {Baluteau} J.-P.,  {Morisset} C.,   {Boisson} C.,  1997, \aap,
  \href {http://adsabs.harvard.edu/abs/1997A%26A...323..217P} {323, 217}

\bibitem[\protect\citeauthoryear{{Pereira}, {Smith}  \& {Cunha}}{{Pereira}
  et~al.}{2005}]{2005A&A...429..993P}
{Pereira} C.~B.,  {Smith} V.~V.,   {Cunha} K.,  2005, \mn@doi [\aap]
  {10.1051/0004-6361:20041020}, \href
  {http://adsabs.harvard.edu/abs/2005A%26A...429..993P} {429, 993}

\bibitem[\protect\citeauthoryear{{Pietrzy{\'n}ski} et~al.,}{{Pietrzy{\'n}ski}
  et~al.}{2013}]{2013Natur.495...76P}
{Pietrzy{\'n}ski} G.,  et~al., 2013, \mn@doi [\nat] {10.1038/nature11878},
  \href {http://adsabs.harvard.edu/abs/2013Natur.495...76P} {495, 76}

\bibitem[\protect\citeauthoryear{{Sanduleak}, {MacConnell}  \&
  {Philip}}{{Sanduleak} et~al.}{1978}]{1978PASP...90..621S}
{Sanduleak} N.,  {MacConnell} D.~J.,   {Philip} A.~G.~D.,  1978, \mn@doi
  [\pasp] {10.1086/130397}, \href
  {http://adsabs.harvard.edu/abs/1978PASP...90..621S} {90, 621}

\bibitem[\protect\citeauthoryear{{Santander-Garc{\'{\i}}a}, {Corradi},
  {Whitelock}, {Munari}, {Mampaso}, {Marang}, {Boffi}  \&
  {Livio}}{{Santander-Garc{\'{\i}}a} et~al.}{2007}]{2007A&A...465..481S}
{Santander-Garc{\'{\i}}a} M.,  {Corradi} R.~L.~M.,  {Whitelock} P.~A.,
  {Munari} U.,  {Mampaso} A.,  {Marang} F.,  {Boffi} F.,   {Livio} M.,  2007,
  \mn@doi [\aap] {10.1051/0004-6361:20065875}, \href
  {http://adsabs.harvard.edu/abs/2007A%26A...465..481S} {465, 481}

\bibitem[\protect\citeauthoryear{{Santander-Garc{\'{\i}}a}, {Corradi},
  {Mampaso}, {Morisset}, {Munari}, {Schirmer}, {Balick}  \&
  {Livio}}{{Santander-Garc{\'{\i}}a} et~al.}{2008}]{2008A&A...485..117S}
{Santander-Garc{\'{\i}}a} M.,  {Corradi} R.~L.~M.,  {Mampaso} A.,  {Morisset}
  C.,  {Munari} U.,  {Schirmer} M.,  {Balick} B.,   {Livio} M.,  2008, \mn@doi
  [\aap] {10.1051/0004-6361:20079212}, \href
  {http://adsabs.harvard.edu/abs/2008A%26A...485..117S} {485, 117}

\bibitem[\protect\citeauthoryear{{Schild} \& {Schmid}}{{Schild} \&
  {Schmid}}{1997}]{1997A&A...324..606S}
{Schild} H.,  {Schmid} H.~M.,  1997, \aap, \href
  {http://adsabs.harvard.edu/abs/1997A%26A...324..606S} {324, 606}

\bibitem[\protect\citeauthoryear{{Schlafly} \& {Finkbeiner}}{{Schlafly} \&
  {Finkbeiner}}{2011}]{2011ApJ...737..103S}
{Schlafly} E.~F.,  {Finkbeiner} D.~P.,  2011, \mn@doi [\apj]
  {10.1088/0004-637X/737/2/103}, \href
  {http://adsabs.harvard.edu/abs/2011ApJ...737..103S} {737, 103}

\bibitem[\protect\citeauthoryear{{Schwarz}}{{Schwarz}}{1991}]{1991A&A...243..469S}
{Schwarz} H.~E.,  1991, \aap, \href
  {http://adsabs.harvard.edu/abs/1991A%26A...243..469S} {243, 469}

\bibitem[\protect\citeauthoryear{{Schwarz} \& {Corradi}}{{Schwarz} \&
  {Corradi}}{1992}]{1992A&A...265L..37S}
{Schwarz} H.~E.,  {Corradi} R.~L.~M.,  1992, \aap, \href
  {http://adsabs.harvard.edu/abs/1992A%26A...265L..37S} {265, L37}

\bibitem[\protect\citeauthoryear{{Shaw}, {Stanghellini}, {Villaver}  \&
  {Mutchler}}{{Shaw} et~al.}{2006}]{2006ApJS..167..201S}
{Shaw} R.~A.,  {Stanghellini} L.,  {Villaver} E.,   {Mutchler} M.,  2006,
  \mn@doi [\apjs] {10.1086/508469}, \href
  {http://adsabs.harvard.edu/abs/2006ApJS..167..201S} {167, 201}

\bibitem[\protect\citeauthoryear{{Shore}, {De Gennaro Aquino}, {Scaringi}  \&
  {van Winckel}}{{Shore} et~al.}{2014}]{2014A&A...570L...4S}
{Shore} S.~N.,  {De Gennaro Aquino} I.,  {Scaringi} S.,   {van Winckel} H.,
  2014, \mn@doi [\aap] {10.1051/0004-6361/201424786}, \href
  {http://adsabs.harvard.edu/abs/2014A%26A...570L...4S} {570, L4}

\bibitem[\protect\citeauthoryear{{Skrutskie} et~al.,}{{Skrutskie}
  et~al.}{2006}]{2006AJ....131.1163S}
{Skrutskie} M.~F.,  et~al., 2006, \mn@doi [\aj] {10.1086/498708}, \href
  {http://adsabs.harvard.edu/abs/2006AJ....131.1163S} {131, 1163}

\bibitem[\protect\citeauthoryear{{Solf}}{{Solf}}{1984}]{1984A&A...139..296S}
{Solf} J.,  1984, \aap, \href
  {http://adsabs.harvard.edu/abs/1984A%26A...139..296S} {139, 296}

\bibitem[\protect\citeauthoryear{{Solf} \& {Ulrich}}{{Solf} \&
  {Ulrich}}{1985}]{1985A&A...148..274S}
{Solf} J.,  {Ulrich} H.,  1985, \aap, \href
  {http://adsabs.harvard.edu/abs/1985A%26A...148..274S} {148, 274}

\bibitem[\protect\citeauthoryear{{Udalski}, {Szyma{\'n}ski}  \&
  {Szyma{\'n}ski}}{{Udalski} et~al.}{2015}]{2015AcA....65....1U}
{Udalski} A.,  {Szyma{\'n}ski} M.~K.,   {Szyma{\'n}ski} G.,  2015, \actaa,
  \href {http://adsabs.harvard.edu/abs/2015AcA....65....1U} {65, 1}

\bibitem[\protect\citeauthoryear{{Ulaczyk} et~al.,}{{Ulaczyk}
  et~al.}{2012}]{2012AcA....62..247U}
{Ulaczyk} K.,  et~al., 2012, \actaa, \href
  {http://cdsads.u-strasbg.fr/abs/2012AcA....62..247U} {62, 247}

\bibitem[\protect\citeauthoryear{{Van Winckel}, {Schwarz}, {Duerbeck}  \&
  {Fuhrmann}}{{Van Winckel} et~al.}{1994}]{1994A&A...285..241V}
{Van Winckel} H.,  {Schwarz} H.~E.,  {Duerbeck} H.~W.,   {Fuhrmann} B.,  1994,
  \aap, \href {http://adsabs.harvard.edu/abs/1994A%26A...285..241V} {285, 241}

\bibitem[\protect\citeauthoryear{{Van Winckel}, {Jorissen}, {Exter}, {Raskin},
  {Prins}, {Perez Padilla}, {Merges}  \& {Pessemier}}{{Van Winckel}
  et~al.}{2014}]{2014A&A...563L..10V}
{Van Winckel} H.,  {Jorissen} A.,  {Exter} K.,  {Raskin} G.,  {Prins} S.,
  {Perez Padilla} J.,  {Merges} F.,   {Pessemier} W.,  2014, \mn@doi [\aap]
  {10.1051/0004-6361/201423650}, \href
  {http://adsabs.harvard.edu/abs/2014A%26A...563L..10V} {563, L10}

\bibitem[\protect\citeauthoryear{{Ventura}, {Stanghellini}, {Dell'Agli},
  {Garc{\'{\i}}a-Hern{\'a}ndez}  \& {Di Criscienzo}}{{Ventura}
  et~al.}{2015}]{2015MNRAS.452.3679V}
{Ventura} P.,  {Stanghellini} L.,  {Dell'Agli} F.,
  {Garc{\'{\i}}a-Hern{\'a}ndez} D.~A.,   {Di Criscienzo} M.,  2015, \mn@doi
  [\mnras] {10.1093/mnras/stv1590}, \href
  {http://adsabs.harvard.edu/abs/2015MNRAS.452.3679V} {452, 3679}

\bibitem[\protect\citeauthoryear{{Woodgate} et~al.,}{{Woodgate}
  et~al.}{1998}]{1998PASP..110.1183W}
{Woodgate} B.~E.,  et~al., 1998, \mn@doi [\pasp] {10.1086/316243}, \href
  {http://adsabs.harvard.edu/abs/1998PASP..110.1183W} {110, 1183}

\bibitem[\protect\citeauthoryear{{Wright} et~al.,}{{Wright}
  et~al.}{2010}]{2010AJ....140.1868W}
{Wright} E.~L.,  et~al., 2010, \mn@doi [\aj] {10.1088/0004-6256/140/6/1868},
  \href {http://adsabs.harvard.edu/abs/2010AJ....140.1868W} {140, 1868}

\makeatother
\end{thebibliography}




\bsp	
\label{lastpage}
\end{document}